\documentclass[12pt]{article}
\usepackage{amsmath}
\usepackage{graphicx}
\numberwithin{equation}{section}

\begin{document}
\baselineskip=24pt
\thispagestyle{empty}
\topskip=0.5cm
\begin{flushright}
\begin{tabular}{c}
September 2006
\end{tabular}
\end{flushright}

\vspace{1cm}

\begin{center}

{\Large\bf Flavor Symmetry and Vacuum Aligned Mass Textures}


\vspace{0.7cm}

Satoru Kaneko$^1$, Hideyuki Sawanaka$^2$, Takaya Shingai$^2$,

Morimitsu Tanimoto$^3$ and Koichi Yoshioka$^4$\\
\vspace{0.2in}

{\sl $^1$ Physics Department, Ochanomizu University, 
Tokyo 112-8610, Japan}\\
{\sl $^2$ Graduate School of Science and Technology,
Niigata University, Niigata 950-2181, Japan}\\
{\sl $^3$ Physics Department, Niigata University, 
Niigata 950-2181, Japan}\\
{\sl $^4$  Physics Department, Kyushu University, 
Fukuoka 812-8581, Japan}\\
\end{center}

\vspace{0.7cm}

\begin{abstract}\noindent
The mass matrix forms of quarks and leptons are discussed in theory
with permutation flavor symmetry. The structure of scalar potential is
analyzed in case that electroweak doublet Higgs fields have 
non-trivial flavor symmetry charges. We find that realistic forms of
mass matrices are obtained dynamically in the vacuum of the theory,
where some of Higgs bosons have vanishing expectation values which
lead to vanishing elements in quark and lepton mass matrices. Mass
textures are realized in the true vacuum and their positions are
controlled by flavor symmetry. An interesting point is that, due to
the flavor group structure, the up and down quark mass matrices are
automatically made different in the vacuum, which lead to
non-vanishing generation mixing. It is also discussed that flavor
symmetry is needed to be broken in order not to have too light
scalars. The lower bounds of Higgs masses are derived from the
experimental data of flavor-changing rare processes such as the
neutral K meson mixing.
\end{abstract}

\newpage
\baselineskip= 24pt

\section{Introduction}

Flavor symmetry is expected to be a clue to understand the masses and
mixing angles of quarks and leptons. It reduces the number of free
parameters in Yukawa couplings, and some testable predictions of
masses and mixing angles generally follow. The discrete non-Abelian
symmetries contain the symmetry groups $S_n$, the dihedral 
groups $D_n$, the binary dihedral (quaternion) groups $Q_n$ and the
tetrahedral groups $A_n$. Some predictive models with discrete flavor
symmetries have been explored by many authors~\cite{S3}-\cite{others}.

The simplest discrete non-Abelian group is $S_3$. The Higgs sector has
a rich structure in theory with the $S_3$ flavor symmetry, e.g.\ there
are two possibilities that $SU(2)$ doublet Higgs fields have trivial
and non-trivial flavor symmetry charges. The former case has been
discussed intensively in the literature (see for example
Ref.~\cite{HY}). The latter case is also expected to have
characteristic phenomenology where Yukawa couplings of quarks and
leptons are described by renormalizable operators.

In this paper, we discuss the mass matrix forms of quarks and leptons
in the case that $SU(2)$ doublet Higgs fields have 
non-trivial $S_3$ flavor charges. We perform the analysis of Higgs
potential at the electroweak scale and examine whether some of Higgs
bosons have vanishing expectation values. In this case, vanishing
elements (texture zeros) of fermion mass matrices are obtained
dynamically in the vacuum of the theory, and their positions are
controlled by flavor symmetry. Such zeros in the mass matrices have
often been assumed by hand~\cite{F3R}-\cite{Rodejohann}, unlike our
scheme. An interesting point of the scheme is that, due to 
the $S_3$ group structure, the up and down quark mass matrices
automatically have different forms at the vacuum, which lead to
non-vanishing generation mixing. The suppression of flavor-changing
neutral currents (FCNC) mediated by multiple Higgs fields will also be
discussed.

This paper is organized as follows. In Section 2, we present some
fundamentals of the $S_3$ group and the $S_3$ symmetry invariant form
of Yukawa couplings (mass matrices) in the supersymmetric case. In
Sections 3 and 4, the symmetry-invariant Higgs potential is
constructed and its structure is analyzed, especially focusing on the
conditions for vanishing vacuum expectation values (VEVs). In 
Section 5, we show that fermion mass matrices favored by the recent
experimental data are highly limited in our $S_3$ framework. In
Sections 6 and 7, several other phenomenological issues, i.e.\ the
Higgs spectrum with flavor symmetry breaking effects and tree-level
FCNC processes, are investigated. Section 8 is devoted to summarizing
our results.

\bigskip
\section{$S_3$ invariant mass matrix}

In this section, the $S_3$-invariant forms of mass matrices are
presented. For more details of the $S_3$ group structure and
symmetry-invariant forms, see, for example, 
Ref.~\cite{HY}. The $S_3$ group has three irreducible representations;
two singlets and one doublet, which we denote throughout this paper 
as ${\bf 1_S}$ (singlet), ${\bf 1_A}$ (pseudo singlet) 
and {\bf 2} (doublet), respectively.
\begin{table}[t]
\begin{center}
\begin{tabular}{c|ccc}
  & ~~${\bf 1_S}$~~ & ~~${\bf 1_A}$~~ & ~~{\bf 2}~~ \\ \hline
${\bf 1_S}$ & ${\bf 1_S}$ & ${\bf 1_A}$ & {\bf 2} \\
${\bf 1_A}$ & ${\bf 1_A}$ & ${\bf 1_S}$ & {\bf 2} \\
{\bf 2} & {\bf 2} & {\bf 2} & ${\bf 2}+{\bf 1_A}+{\bf 1_S}$ 
\end{tabular}
\caption{$S_3$ tensor products.}
\label{tp}
\end{center}
\end{table}
The decomposition of tensor products is shown in Table~\ref{tp}. A
non-trivial product 
is ${\bf 2}\times{\bf 2}={\bf 2}+{\bf 1_A}+{\bf 1_S}$. For two
doublets $\psi=(\psi_1,\psi_2)^T$ and $\phi=(\phi_1,\phi_2)^T$, that
is explicitly given by
\begin{eqnarray}
 \psi\times\phi &=&
 \begin{pmatrix}
   \psi^\dag_1 \phi_2 \\ \psi^\dag_2 \phi_1
 \end{pmatrix}_{\bf 2}
 +(\psi^\dag_1 \phi_1-\psi^\dag_2 \phi_2)_{\bf 1_A}
 +(\psi^\dag_1 \phi_1+\psi^\dag_2 \phi_2)_{\bf 1_S}\ ,
\end{eqnarray}
where the suffixes in the right hand side of the equation 
denote $S_3$ representations. Another form of product is written by
using $\psi_{C}\equiv (\psi^\ast_2,\psi^\ast_1)^T$ which behaves as
a doublet. The tensor product of $\psi_{C}$ with another 
doublet $\phi$ becomes
\begin{eqnarray}
 \psi_{C} \times \phi &=&
 \begin{pmatrix}
   \psi_2\phi_2 \\ \psi_1\phi_1
 \end{pmatrix}_{\bf 2}
 +(\psi_1\phi_2-\psi_2\phi_1)_{\bf 1_A}
 +(\psi_1\phi_2+\psi_2\phi_1)_{\bf 1_S}\ .
\label{tensor}
\end{eqnarray}
Since this latter form of product does not contain any complex
conjugates, it is useful for describing, e.g.\ Majorana masses for
neutrinos and holomorphic terms in supersymmetric theory. It is
mentioned that the two types of $S_3$ invariants given above
correspond to two invariant tensors of $SU(2)$, which 
contains $S_3$ as a subgroup.

Let us discuss quark/lepton mass matrices, where both left-handed and
right-handed fermions transform under a single $S_3$ symmetry. We
suppose that two of three generations belong to $S_3$ doublets and the
others are singlets. In this paper, we consider supersymmetric theory 
and then matter superfields of first two 
generations $\Psi_{L_{1,2}}$ and $\Psi_{R_{1,2}}$ are treated 
as $S_3$ doublets, into which left and right-handed fermions are
embedded as $\psi_{L_{1,2}}$ and $\psi^{\,c}_{R_{1,2}}$, respectively.
As for $SU(2)$ weak doublet Higgses, all three types 
of $S_3$ irreducible representations are introduced; a 
doublet $(H_1,H_2)^T$, a pseudo singlet $H_A$ and a 
singlet $H_S$ for each of up and down type Higgs sector. It is found
from the tensor product (\ref{tensor}) that the most general
supersymmetric Yukawa (mass) terms are written as
\begin{equation}
  W_y \,=\, \Psi_L{}_i\,(M_D){}_{ij}\,\Psi_R{}_j\ ,
\end{equation}
\begin{equation}
  M_D \,=\, \left(
  \begin{tabular}{cc|c}
    $aH_{1}$ & $bH_{S}+cH_{A}$ & $dH_{2}$ \\
    $bH_{S}-cH_{A}$ & $aH_{2}$ & $dH_{1}$ \\ \hline 
    $eH_{2}$ & $eH_{1}$ & $fH_{S}$        
  \end{tabular} \right),
  \label{MD}
\end{equation}
where $a$, $b$, $\cdots$, $f$ are independent Yukawa coupling constants.

It is noticed that, if some of Higgs bosons have vanishing expectation
values in the vacuum of the theory, corresponding mass matrix elements
vanish and then mass (not Yukawa) texture is realized. That is the
scheme we adopt in this paper for quark/lepton mass matrices.
A non-trivial issue is whether such VEVs are obtained in the
electroweak symmetry breaking vacuum. Therefore the Higgs potential should
be carefully examined.

Finally, the $S_3$-invariant bare Majorana mass for matter 
superfield $\Psi_i$ ($i=1,2,3$) is given by
\begin{equation}
  W_m \,=\, \Psi_i\,(M_R)_{ij}\,\Psi{}_j\ ,
\end{equation}
\begin{equation}
  M_R \,=\, \left(
  \begin{tabular}{cc|c}
    & $M_1$ & \\
    $M_1$ & & \\ \hline
    & & $M_2$
  \end{tabular} \right).
\end{equation}

\bigskip
\section{$S_3$ Higgs scalar potential}

Let us consider supersymmetric theory with $S_3$ flavor symmetry and 
introduce the following Higgs superfields:
\begin{eqnarray*}
\left\{ \begin{array}{ccl}
  \hat{H}_{uS},\ \hat{H}_{dS} &:& S_3 \ {\rm singlets} \\
  \hat{H}_{uA},\ \hat{H}_{dA} &:& S_3 \ {\rm pseudo\ singlets} \\
  (\hat{H}_{u1},\hat{H}_{u2}),\ (\hat{H}_{d1},\hat{H}_{d2}) 
  &:& S_3 \ {\rm doublets}
\end{array}\right.\ .
\end{eqnarray*}
The symmetry-invariant supersymmetric Lagrangian is
\begin{equation}
  {\cal L} \,=\, \int d^2\theta d^2\bar\theta\ K 
  +\left(\int d^2\theta\ W + {\rm h.c.}\right),
\end{equation}
\begin{eqnarray}
  K &=& \!\!\sum_{\alpha = S, A, 1, 2}
  \left( \hat{H}^\dagger_{u\alpha} e^{G_u} \hat{H}_{u\alpha} 
    +\hat{H}^\dagger_{d\alpha} e^{G_d} \hat{H}_{d\alpha} \right), \\ 
  W &=& \mu_S \hat{H}_{uS} \epsilon \hat{H}_{dS} 
  +\mu_A \hat{H}_{uA} \epsilon \hat{H}_{dA} 
  +\mu_D \left( \hat{H}_{u1} \epsilon \hat{H}_{d2} 
    +\hat{H}_{u2} \epsilon \hat{H}_{d1} \right) ,
\end{eqnarray}
where $G_{u,d}\equiv\pm\frac{1}{2}g_{Y}\hat{V}_{Y}+g_2\hat{V}_2$ with
the electroweak vector superfields $\hat{V}_Y$ and $\hat{V}_2$, 
and $g_Y$ and $g_2$ are the corresponding gauge coupling
constants. The mass parameters $\mu_{S,A,D}$ are generally complex 
and $\epsilon$ is the antisymmetric tensor for $SU(2)$ weak 
indices ($\epsilon^{12}=1$). The Higgs scalar potential is then given by
\begin{equation}
  V \,=\, V_{\rm susy} + V_{\rm soft}\ ,
\end{equation}
\begin{eqnarray}
  V_{\rm susy} &=& \frac{1}{2} D^2_Y +\frac{1}{2}\sum_{a=1,2,3}(D^a)^2 
  \nonumber \\
  && +\left|\mu_S\right|^2 \left( H^\dagger_{uS}H_{uS} 
    +H^\dagger_{dS}H_{dS} \right) 
  +\left|\mu_A\right|^2 \left( H^\dagger_{uA}H_{uA} 
    +H^\dagger_{dA}H_{dA} \right) \nonumber \\
  && +\left|\mu_D\right|^2 \left( H^\dagger_{u1}H_{u1} 
    +H^\dagger_{d1}H_{d1} +H^\dagger_{u2}H_{u2} 
    +H^\dagger_{d2}H_{d2}\right), 
\end{eqnarray}
\begin{eqnarray}
  V_{\rm soft} &=& m^2_{uS}H^\dagger_{uS}H_{uS}
  +m^2_{dS} H^\dagger_{dS}H_{dS} +\left(b_S H_{uS}\epsilon H_{dS} 
    +{\rm h.c.} \right)  \nonumber \\
  && +m^2_{uA} H^\dagger_{uA}H_{uA} +m^2_{dA} H^\dagger_{dA}H_{dA} 
  +\left( b_A H_{uA}\epsilon H_{dA} +{\rm h.c.} \right) \nonumber \\
  && +m^2_{uD} (H^\dagger_{u1}H_{u1} +H^\dagger_{u2}H_{u2})
  +m^2_{dD} (H^\dagger_{d1}H_{d1} +H^\dagger_{d2}H_{d2})  \nonumber \\
  && +\left[ b_D \left( H_{u1}\epsilon H_{d2} +H_{u2}\epsilon H_{d1}
    \right) +{\rm h.c.} \right]\ ,
\end{eqnarray}
where $b_{x}$, $m_{ux}$ and $m_{dx}$ ($x=S,A,D$) are 
the holomorphic and non-holomorphic mass parameters of supersymmetry breaking, respectively.
The $D$ terms are explicitly given by
\begin{eqnarray}
  D_Y &=& -\frac{1}{2} g_Y\left( H^\dagger_{uS}H_{uS} 
    -H^\dagger_{dS}H_{dS} +H^\dagger_{uA}H_{uA} -H^\dagger_{dA}H_{dA}
  \right. \nonumber \\
  && +\left. H^\dagger_{u1}H_{u1} -H^\dagger_{d1}H_{d1}
    +H^\dagger_{u2}H_{u2} -H^\dagger_{d2}H_{d2} \right), \\
  D^a &=& -g_2 \left( H^\dagger_{uS} T^a H_{uS} 
    +H^\dagger_{dS} T^a H_{dS} +H^\dagger_{uA} T^a H_{uA} 
    +H^\dagger_{dA} T^a H_{dA} \right. \nonumber \\
  && +\left. H^\dagger_{u1} T^a H_{u1} 
    +H^\dagger_{d1} T^a H_{d1} +H^\dagger_{u2} T^a H_{u2}
    +H^\dagger_{d2} T^a H_{d2} \right) ,
\end{eqnarray}
with $T^a$ ($a=1,2,3$) is the $SU(2)$ weak generators. It should be
noted that $V_{\rm soft}$ is introduced as the most general
supersymmetry-breaking Higgs soft terms compatible with 
the $S_3$ symmetry.

We here mention a possibility of spontaneous breakdown of CP
symmetry. The following terms in $V_{\rm soft}$ which contain complex
couplings could be sources of CP violation:
\begin{equation}
  b_S H_{uS}\epsilon H_{dS} +b_A H_{uA}\epsilon H_{dA}
  +b_D \left( H_{u1}\epsilon H_{d2} +H_{u2}\epsilon H_{d1} \right) 
  +{\rm h.c.} \ .
\end{equation}
Analyzing the stationary conditions for the Higgs potential, it is
understood that the phases of Higgs VEVs should 
satisfy $\theta_{u\alpha}+\theta_{d\alpha}=0$ ($\alpha=S,A,1,2$) 
where $\theta_{u\alpha(d\alpha)}$ are the up(down)-type Higgs phases.
While these phases do not directly appear in the Higgs scalar
potential at the vacuum, they cannot be removed in general by field
redefinitions unlike in the minimal supersymmetric standard 
model (MSSM)\@. Therefore there is a possibility of causing
spontaneous CP breakdown by taking account of complex phases of Higgs
VEVs in Higgs-mediated processes.

\bigskip
\section{Analysis of Higgs scalar potential}

Hereafter we assume that the charged Higgs fields do not develop
nonzero VEVs so as to make the $U(1)_{EM}$ symmetry remains intact
after the electroweak symmetry breaking. The $S_3$ Higgs scalar
potential we analyze thus becomes
\begin{eqnarray}
  V &=& \left( \left|\mu_S\right|^2 +m^2_{uS} \right)v_{uS}^2 
  +\left( \left|\mu_S\right|^2 +m^2_{dS} \right) v_{dS}^2 
  -2b_Sv_{uS}v_{dS} \nonumber \\ 
  && +\left( \left|\mu_A\right|^2 +m^2_{uA} \right) v_{uA}^2 
  +\left( \left|\mu_A\right|^2 +m^2_{dA} \right) v_{dA}^2 
  -2b_Av_{uA}v_{dA} \nonumber \\
  && +\left( \left|\mu_D\right|^2 +m^2_{uD} \right)
  \left( v_{u1}^2 + v_{u2}^2 \right) 
  +\left( \left|\mu_D\right|^2 +m^2_{dD} \right)
  \left( v_{d1}^2 + v_{d2}^2 \right)  \nonumber \\[1mm]
  && -2b_D \left( v_{u1}v_{d2} + v_{u2}v_{d1} \right) 
  \,+\frac{g^2_Y+g^2_2}{8} X^2,
  \label{potential}
\end{eqnarray}
\begin{equation}
  X \,\equiv\, v^2_{uS}-v^2_{dS}+v^2_{uA}-v^2_{dA}+v^2_{u1}-v^2_{d1}
  +v^2_{u2}-v^2_{d2},
\end{equation}
where $v_x$ are the absolute values of Higgs scalars:
\begin{eqnarray}&&
  v_{uS} \,=\, \left|\left\langle H^0_{uS}\right\rangle\right|, \quad
  v_{uA} \,=\, \left|\left\langle H^0_{uA}\right\rangle\right|, \quad
  v_{u1} \,=\, \left|\left\langle H^0_{u1}\right\rangle\right|, \quad
  v_{u2} \,=\, \left|\left\langle H^0_{u2}\right\rangle\right|, 
  \nonumber \\ &&
  v_{dS} \,=\, \left|\left\langle H^0_{dS}\right\rangle\right|, \quad
  v_{dA} \,=\, \left|\left\langle H^0_{dA}\right\rangle\right|, \quad
  v_{d1} \,=\, \left|\left\langle H^0_{d1}\right\rangle\right|, \quad
  v_{d2} \,=\, \left|\left\langle H^0_{d2}\right\rangle\right|,
\end{eqnarray}
and the parameters $b_{S,A,D}$ have been chosen to be real positive 
by field redefinitions. Notice that the scalar 
potential (\ref{potential}) is invariant under the label-exchanging
transformations, $1\leftrightarrow 2$, $S\leftrightarrow A$ 
and/or $u\leftrightarrow d$.

First we discuss the instability of scalar potential at the origin of
field space. The potential parameters have to satisfy at least one of
the following conditions in order to break the electroweak gauge
symmetry:
\begin{eqnarray}
  b_S^2 &>& (\left|\mu_S\right|^2 +m^2_{uS})(\left|\mu_S\right|^2
  +m^2_{dS})\ , \\
  b_A^2 &>& (\left|\mu_A\right|^2 +m^2_{uA})(\left|\mu_A\right|^2 
  +m^2_{dA})\ , \\
  b^2_D &>& (\left|\mu_D\right|^2 +m^2_{uD})(\left|\mu_D\right|^2 
  +m^2_{dD})\ .
\end{eqnarray}
Then the equations of motion become
\begin{eqnarray} &&
  (\left|\mu_S\right|^2 +m^2_{uS})v_{uS} \,=\, 
  b_S v_{dS} -X^\prime v_{uS}\ , \quad
  (\left|\mu_S\right|^2 +m^2_{dS})v_{dS} \,=\,
  b_S v_{uS} +X^\prime v_{dS}\ , 
  \label{equation1} \\ &&
  (\left|\mu_A\right|^2 +m^2_{uA})v_{uA} \,=\, 
  b_A v_{dA} -X^\prime v_{uA}\ , \quad
  (\left|\mu_A\right|^2 +m^2_{dA})v_{dA} \,=\,
  b_A v_{uA} +X^\prime v_{dA}\ , 
  \label{equation2} \\ &&
  (\left|\mu_D\right|^2 +m^2_{uD})v_{u1} \,=\,
  b_D v_{d2} -X^\prime v_{u1}\ , \quad
  (\left|\mu_D\right|^2 +m^2_{dD})v_{d2} \,=\,
  b_D v_{u1} +X^\prime v_{d2}\ ,
  \label{equation3} \\ &&
  (\left|\mu_D\right|^2 +m^2_{uD})v_{u2} \,=\,
  b_D v_{d1} -X^\prime v_{u2}\ , \quad
  (\left|\mu_D\right|^2 +m^2_{dD})v_{d1} \,=\,
  b_D v_{u2} + A^\prime v_{d1}\ ,
  \label{equation4}
\end{eqnarray}
where $X^\prime=\frac{g^2_Y+g^2_2}{4}X$. These equations
(\ref{equation1})-(\ref{equation4}) depend respectively 
on $v_{uS(dS)}$, $v_{uA(dA)}$, $v_{u1(d2)}$ and $v_{u2(d1)}$, except
in the $X'$ parts. The $Z$ boson mass is given by
\begin{equation}
  m^2_Z \,=\, \frac{1}{2} (g^2_Y+g^2_2) 
  \left( v^2_{uS}+v^2_{dS}+v^2_{uA}+v^2_{dA}
    +v^2_{u1}+v^2_{d1}+v^2_{u2}+v^2_{d2} \right).
\end{equation}
In addition, substituting (\ref{equation1})-(\ref{equation4}) for the
scalar potential, we find the depth of the potential at the minimum:
\begin{equation}
  V_{\rm min} \,=\, -\frac{g^2_Y+g^2_2}{8}X^2 \,\leq\, 0\ .
  \label{depth}
\end{equation}
This expression also holds in the case of $S_3$ symmetry breaking
model discussed in later section.

Next we study the possibility of having vanishing VEVs in the minimum
of the potential. In general, (\ref{equation1})-(\ref{equation4}) are
the coupled equations through the $X$ parts. We separate these
equations into three parts  for the singlet (\ref{equation1}), the
pseudo singlet (\ref{equation2}), and the doublet (\ref{equation3})
and (\ref{equation4}). As for the singlet and pseudo singlet parts, 
each coefficients of $X'$ in (\ref{equation1}) and (\ref{equation2})
are $v_{u(d)S}$ and $v_{u(d)A}$, respectively. Therefore vanishing
VEVs makes the equations trivial within each sector. There is however
a bit difference in the $S_3$ doublet part [Eqs.~(\ref{equation3}) and
(\ref{equation4})], that is, they cannot be separated from each
other. These equations have not only a relation through the $X$ part
but also a common parameter $b_D$ which originates from the symmetry
invariance.

\subsection{$S_3$ singlet Higgses}

First let us examine the vanishing VEVs of $S_3$ singlet Higgs
fields. In this case, we take the other Higgs VEVs arbitrary. The
analysis is the same for pseudo singlet Higgs fields. There are two
possible patterns of zero VEVs; one is that both of the up and
down-type Higgs VEVs are zero and another is that only one of them is
zero.

\begin{itemize}
\item $v_{uS}=v_{dS}=0$

This solution always exists. Though such a solution does not make
sense in models with one pair of Higgs doublets like the MSSM, it is
possible in the present case as long as some of instability conditions
for other Higgs parts are satisfied.

\item $v_{uS}=0$, $v_{dS}\neq0$ ~~or~~ $v_{dS}=0$, $v_{uS}\neq0$

It is found that the parameter condition $b_S=0$ is necessary for this
solution. For example, if one supposes $v_{uS}=0$ and $v_{dS}\neq 0$,
the stationary conditions mean
\begin{equation}
  b_S \,=\, 0 \quad {\rm ~and~} \quad
  \left|\mu_S\right|^2 +m^2_{dS} -\frac{g^2_Y+g^2_2}{4}
  \left(-v^2_{dS}+X_0\right) \,=\, 0,
\end{equation}
where we have defined $X_0=X|_{v_{uS}=v_{dS}=0}$. Therefore as long as
the parameter $b_S$ is nonzero, the solution with one of VEVs being
zero does not exist. It is noted that the condition $b_S=0$ is
difficult to be satisfied exactly at any scale, and hence such a type
of solution may not be realistic.
\end{itemize}

\subsection{$S_3$ doublet Higgses}
\label{DHiggs}

Including $S_3$ doublet Higgs fields makes the theory complex but is
promising. Unlike the case of $S_3$ singlet Higgses, all the four
VEVs, $v_{u1}$, $v_{u2}$, $v_{d1}$ and $v_{d2}$ should be
simultaneously taken into account because $b_D$ is a common parameter
as we mentioned above. In the analysis, we take the singlet Higgs
VEVs, $v_{uS}$, $v_{dS}$, $v_{uA}$ and $v_{dA}$, to be arbitrary. It is
important to notice that the stationary conditions for $S_3$ doublet
Higgses are described (except in the $X$ parts) by two pairs of VEVs:
Eq.~(\ref{equation3}) for $(v_{u1},v_{d2})$ and
Eq.~(\ref{equation4}) for $(v_{u2},v_{d1})$. It can be seen from the
result in the previous section that the existence of VEV pairs
means that the VEV forms such as $v_{u1}=0$ and $v_{d2}\neq0$ are not
allowed, i.e.\ only one VEV in each pair does not develop a nonzero
VEV unless an unlikely condition $b_D=0$ is satisfied.

There are 16 ($=2^4$) patterns for the 4 VEVs of $S_3$ doublets. Among
them, only 7 patterns are theoretically independent due to the
label-exchanging invariances of the potential: $1\leftrightarrow 2$
and $u\leftrightarrow d$, mentioned before. For example, in the case
that $v_{u1}=v_{d2}=0$ and $v_{u2},\,v_{d1}\neq0$, the equations of
motion give
\begin{equation}
  2|\mu_D|^2 +m^2_{uD} +m^2_{dD} \,=\,
  b_D\left( \frac{v_{u2}}{v_{d1}}+\frac{v_{d1}}{v_{u2}} \right).
\end{equation}
A similar form of vacuum equation is obtained for a theoretically
equivalent case that $v_{u2}=v_{d1}=0$ and $v_{u1},\,v_{d2}\neq0$ by
use of label exchanges. Therefore it is enough to consider either of
these patterns. All 7 possible patterns obtained from the potential
analysis are shown in Table~\ref{doublet}.
\begin{table}[t]
\begin{center}
\begin{tabular}{|c|c|c|c||c|} \hline
$ v_{u1}$ & $v_{u2}$ & $v_{d1}$ & $v_{d2}$ & conditions \\ \hline\hline
$0$ & $0$ & $0$ & $0$ & \\ \hline
$0$ & $0$ & $\not\!0$ & $0$ & $b_D=0$ \\ \hline
$0$ & $\not\!0$ & $\not\!0$ & $0$ & \\ \hline
$0$ & $0$ & $\not\!0$ & $\not\!0$ & $b_D=0$ \\ \hline
$0$ & $\not\!0$ & $0$ & $\not\!0$ & 
$b_D=0$, ~ $2\left|\mu_D\right|^2 +m^2_{uD} +m^2_{dD}=0$ \\ \hline
$\not\!0$ & $\not\!0$ & $0$ & $\not\!0$ & 
$b_D=0$, ~ $2\left|\mu_D\right|^2 +m^2_{uD} +m^2_{dD}=0$ \\ \hline
$\not\!0$ & $\not\!0$ & $\not\!0$ & $\not\!0$ & \\ \hline
\end{tabular}
\end{center}
\caption{All representative VEV patterns of the $S_3$ doublet Higgs
fields. The blank entries denote no needs of parameter
conditions.\bigskip}
\label{doublet} 
\end{table}
It is noted that the parameter conditions given in Table~\ref{doublet}, 
i.e.\ $b_D=0$ and $2|\mu_D|^2+m^2_{uD}+m^2_{dD}=0$, are unlikely to be
satisfied exactly, because these relations are not protected by
symmetry arguments and generally unstable against quantum corrections
(though there are some models for a vanishing $b$ parameter at a given
scale~\cite{bzero}). Therefore in this paper we do not consider the
solutions with non-trivial parameter conditions. In the end, we have
found the relevant vacuum solutions without parameter conditions,
which are presented in Table~\ref{doublet2}.
\begin{table}[t]
\begin{center}
\begin{tabular}{|c|c|c|c|} \hline
$v_{u1}$ & $v_{u2}$ & $v_{d1}$ & $v_{d2}$ \\ \hline\hline
$0$ & $0$ & $0$ & $0$ \\ \hline
$0$ & $\not\!0$ & $\not\!0$ & $0$ \\ \hline
$\not\!0$ & $0$ & $0$ & $\not\!0$ \\ \hline
$\not\!0$ & $\not\!0$ & $\not\!0$ & $\not\!0$ \\ \hline
\end{tabular}
\end{center}
\caption{All VEV patterns of the $S_3$ doublet Higgs fields without
parameter conditions.\bigskip}
\label{doublet2}
\end{table}

To summarize the results in Section 4, we have found all possible
minima of the scalar potential for $S_3$ singlet and doublet Higgs
fields (Table~\ref{summary}). These solutions do not require any exact
tuning of Lagrangian parameters for electroweak symmetry breaking and
are physically available. The potential depth at each minimum is
controlled by Higgs mass parameters in the Lagrangian, and any of the
VEV patterns in Table~\ref{summary} can be made the global vacuum of
the theory, as seen in Eq.~(\ref{depth}).
\begin{table}[t]
\begin{center}
\begin{tabular}{|c|c|c|c|c|c|c|c|} \hline
$v_{uS}$ & $v_{dS}$ & $v_{uA}$ & $v_{dA}$ & $v_{u1}$ & $v_{u2}$ 
& $v_{d1}$ & $v_{d2}$ \\ \hline \hline
 $0$ & $0$ & $0$ & $0$ & $0$ &     &     & $0$ \\ \hline
 $0$ & $0$ & $0$ & $0$ &     & $0$ & $0$ &     \\ \hline
 $0$ & $0$ & $0$ & $0$ &     &     &     &     \\ \hline
 $0$ & $0$ &     &     & $0$ & $0$ & $0$ & $0$ \\ \hline
 $0$ & $0$ &     &     & $0$ &     &     & $0$ \\ \hline
 $0$ & $0$ &     &     &     & $0$ & $0$ &     \\ \hline
 $0$ & $0$ &     &     &     &     &     &     \\ \hline
     &     & $0$ & $0$ & $0$ & $0$ & $0$ & $0$ \\ \hline
     &     & $0$ & $0$ & $0$ &     &     & $0$ \\ \hline
     &     & $0$ & $0$ &     & $0$ & $0$ &     \\ \hline
     &     & $0$ & $0$ &     &     &     &     \\ \hline
     &     &     &     & $0$ & $0$ & $0$ & $0$ \\ \hline
     &     &     &     & $0$ &     &     & $0$ \\ \hline
     &     &     &     &     & $0$ & $0$ &     \\ \hline
     &     &     &     &     &     &     &     \\ \hline
\end{tabular}
\end{center}
\caption{All possible minima of the scalar potential for $S_3$ singlet
and doublet Higgs fields without tuning of Lagrangian parameters for
electroweak symmetry breaking. The blank entries denote
non-vanishing VEVs.\bigskip}
\label{summary}
\end{table}

\bigskip
\section{Quark and lepton mass textures} 

As a phenomenological application of the vacuum analysis performed in
the previous section, we study in this section the mass matrix forms
of quarks and leptons derived from $S_3$ flavor symmetry. Let us
consider a supersymmetric theory with three-generation quark and
lepton superfields in addition to the Higgs content previously
analyzed. The three generations belong to a 
reducible {\bf 3} representation of $S_3$, that is, two of three
generations behave as a doublet and the other is a singlet or a pseudo
singlet. When the first two generation superfields consist 
of $S_3$ doublets, the most general form of mass matrix is given 
by (\ref{MD}) with non-vanishing Higgs VEVs.

A usual approach often seen in the literature to obtain realistic
forms of mass matrices is to control coupling constants of effective
Yukawa operators with additional symmetries or to adjust coupling
constants by hand so that the experimental data is reproduced. Unlike
these approaches, our strategy in this paper is to dynamically realize
mass matrix forms (mass textures) in the vacuum of the theory. That
is, as discussed in the previous section, some of Higgs VEVs vanish at
the minimum of scalar potential, which in turn lead to 
texture (zero) forms of quark/lepton mass matrices. The available
candidates of vacua are listed in Table~\ref{summary}. One can see
from the table that the VEV structures of up and down-type Higgs
fields are not parallel and rather different due to the group
properties of $S_3$ flavor symmetry. This fact is favorable in light
of the experimental data of fermion masses and mixing angles, which
data is well known to show that the up and down quark sectors would
have highly different generation structures.

In the following, we assume as an example that 
the $S_3$ representations of three-generation matter fields 
are ${\bf 2}+{\bf 1_S}$. If one adopts ${\bf 1_A}$ instead 
of ${\bf 1_S}$, phenomenological results are changed according to
Eq.~(\ref{tensor}). Examining the vacuum patterns in
Table~\ref{summary}, we find that the following case with four zero
VEVs leads to the simplest texture (i.e.\ the maximal number of zero
matrix elements) with non-trivial flavor mixing:
\begin{equation}
  v_{u1} = v_{d2} = v_{uS} = v_{dS} = 0 , \qquad
  v_{u2},\; v_{d1},\; v_{uA},\; v_{dA} \,\neq\, 0.
  \label{vev0}
\end{equation}
The VEV pattern obtained by interchanging the 
labels $1\leftrightarrow2$ is also the case. That can be covered by
exhausting the generation label exchanges of matter fields, and
therefore we safely focus on the vacuum (\ref{vev0}) and examine all
types of $S_3$ charge assignments of matter fields. In what follows,
we use the notation 
that $Q_i$, $u_i$, $d_i$, $L_i$, $e_i$, and $\nu_i$ ($i=1,2,3$) are
the superfields of left-handed quarks, right-handed up quarks,
right-handed down quarks, left-handed leptons, right-handed charged
leptons and right-handed neutrinos, respectively.

\subsection{Quark mass textures}

When the first and second generation quarks behave as $S_3$ doublets, 
the up and down quark mass textures in the vacuum (\ref{vev0}) are
read from the generic form of mass matrix (\ref{MD}):
\begin{equation}
  M_u \,=\, \left(
    \begin{array}{ccc}
      & \beta v_{uA} & \gamma v_{u2} \\
      -\beta v_{uA} & \alpha v_{u2} & \\
      \delta v_{u2} & & 
    \end{array}\right) \ , \qquad
  M_d \,=\, \left(
    \begin{array}{ccc}
      \bar\alpha v_{d1} & \bar\beta v_{dA} & \\
      -\bar\beta v_{dA} & & \bar\gamma v_{d1} \\
      & \bar\delta v_{d1} & 
    \end{array}\right) \ ,
\end{equation}
where the blank entries mean zeros, 
and $\alpha$, $\beta$, $\cdots$, $\bar \gamma$ and $\bar \delta$ are
the Yukawa coupling constants. In this flavor basis, the above
matrices appear not to lead to appropriate mass hierarchies and
generation mixing. We have exhausted the $S_3$ charge assignments of
matter fields and found that only the following 4 cases are almost
consistent with the current experimental data of fermion masses and
mixing angles.

\subsubsection{$\bf (Q_2,Q_1)+Q_3$, $\;\;\bf (u_3,u_2)+u_1$, 
$\;\;\bf (d_2,d_3)+d_1$}

The title of this subsection means the $S_3$ charge assignment that
the second and first generations of left-handed quark superfields
consist of a doublet and the third one $Q_3$ is a singlet, and
similarly for $u_i$ and $d_i$. In this case, the quark mass matrices
are given by
\begin{equation}
  M_u \,=\, \left(
  \begin{array}{ccc}
    & b_u & -c_u \\ d_u & c_u & \\ & & i_u
  \end{array}\right) \ , \qquad\quad
  M_d \,=\, \left(
  \begin{array}{ccc}
    a_d & -b_d & \\ & e_d & b_d \\ & & i_d
  \end{array}\right) \ .
\end{equation}
In the first order approximation, they predict the quark mass
eigenvalues and mixing angles
\begin{eqnarray}
 & m_u \,=\, -\frac{b_ud_u}{c_u} \ ,\qquad\quad 
  m_c \,=\, c_u \ ,  \qquad\quad 
  m_t \,=\, i_u \ , \\
 & m_d \,=\, a_d \ , \qquad\quad 
  m_s \,=\, e_d \ , \qquad\quad 
  m_b \,=\, i_d \ ,    \\ 
 & V_{us} \,=\, -\frac{b_u}{c_u}-\frac{b_d}{e_d} \ , \qquad\quad 
  V_{cb} \,=\, \frac{b_d}{i_d} \ , \qquad\quad
  V_{ub} \,=\, \frac{c_u}{i_u}-\frac{b_ub_d}{c_ui_d} \ ,
\end{eqnarray}
which lead to a relation among the observables:
\begin{equation}
  V_{ub} -\frac{m_c}{m_t} \,=\, 
  V_{cb}\left(V_{us}+\frac{m_b}{m_s}V_{cb}\right) \ .
  \label{rel1}
\end{equation}
Such parameter-independent relation may be useful to examine whether
the model can well describe the observations. If one evaluates the
relation (\ref{rel1}) with respect to the mixing matrix 
element $V_{cb}$, one third of the observed value is reproduced, which
implies a necessity of some modification of the mass matrix forms.

\subsubsection{$\bf (Q_3,Q_2)+Q_1$, $\;\;\bf (u_1,u_2)+u_3$, 
$\;\;\bf (d_3,d_2)+d_1$}
 
In the second case, $(Q_3,Q_2)$, $(u_1,u_2)$ and $(d_3,d_2)$ are
the $S_3$ doublets. The mass matrices then take the forms:
\begin{equation}
  M_u \,=\, \left(
  \begin{array}{ccc}
    a_u & & \\ -d_u & e_u & \\ & d_u & i_u
  \end{array}\right) \ , \qquad\quad
  M_d \,=\, \left(
  \begin{array}{ccc}
    & b_d & \\ d_d & & -f_d \\ & f_d & i_d
  \end{array}\right) \ .
\end{equation}
The flavor charge assignments of $Q_i$ and $d_i$ are the same, which
leads to a generation-symmetric texture form like the Fritzsch
ansatz~\cite{F3R} for the down quark mass matrix. Also such charge
assignment might be relevant to the flipped $SU(5)$ unified
models~\cite{flipped}. The prediction for the quark mass eigenvalues
and mixing angles is given by
\begin{eqnarray}
  & m_u \,=\, \frac{a_ue_u}{\sqrt{d_u^2+e_u^2}} \ , \qquad\quad 
    m_c,=\, \sqrt{d_u^2+e_u^2} \ , \qquad\quad 
    m_t \,=\, i_u \ , \\
  & m_d \,=\, -\frac{b_dd_di_d}{f_d^2} \ ,\qquad\quad 
    m_s \,=\, \frac{f_d^2}{i_d} \ ,\qquad\quad 
    m_b \,=\, i_d \ , \\
  & V_{us} \,=\, \frac{b_di_d}{f_d^2} \ , \qquad\quad 
    V_{cb} \,=\, -\frac{f_d}{i_d} \ ,\qquad\quad
    V_{ub} \,=\, \frac{b_df_d}{i_d^2} \ ,
\end{eqnarray}
in the first order approximation. It is noted that the leading
contributions to the mixing angles mainly come from the down quark
sector, and then satisfy the following two relations among the
observables:
\begin{equation}
  |V_{ub}| \,=\, |V_{us}|\left(\frac{m_s}{m_b}\right)^{3/2}, \qquad\quad
  |V_{cb}| \,=\, \sqrt{\frac{m_s}{m_b}} \ .
\end{equation}
These do not so largely deviate from the experimental data, but
predict a bit large (small) value for the mixing matrix element
$V_{cb}$ (for the mass eigenvalue $m_s$).

\subsubsection{$\bf (Q_2,Q_3)+Q_1$, $\;\;\bf (u_2,u_3)+u_1$,
$\;\;\bf (d_2,d_1)+d_3$} 

The third pattern is the case that $(Q_2,Q_3)$, $(u_2,u_3)$ 
and $(d_2,d_1)$ are the $S_3$ doublets, which lead to the mass
textures:
\begin{equation}
  M_u \,=\, \left(
  \begin{array}{ccc}
    & b_u & \\ d_u & & f_u \\ & -f_u & i_u
  \end{array}\right)  \ , \qquad\quad
  M_d \,=\, \left(
  \begin{array}{ccc}
    a_d & & \\ d_d & e_d & \\ & -d_d & i_d 
  \end{array}\right) \ .
\end{equation}
This type of charge assignment is compatible with $SU(5)$ grand
unification~\cite{su5}. The masses and mixing angles are determined in
the first order as
\begin{eqnarray}
 & m_u \,=\, -\frac{b_ud_ui_u}{f_u^2} \ , \qquad
  m_c \,=\, \frac{f_u^2}{i_u} \ , \qquad\quad 
  m_t \,=\, i_u \ , \\
 &  m_d \,=\, \frac{a_de_d}{\sqrt{d_d^2+e_d^2}} \ , \qquad\quad 
  m_s \,=\, \sqrt{d_d^2+e_d^2} \ , \qquad\quad 
  m_b \,=\, i_d \ , \\
 &  V_{us} \,=\, -\frac{b_ui_u}{f_u^2} \ , \qquad\quad 
  V_{cb} \,=\, -\frac{f_u}{i_u}\ ,  \qquad\quad
  V_{ub} \,=\, \frac{b_u}{f_u} \ .
\end{eqnarray}
It is noted that the leading contributions to the mixing angles mainly
come from the up quark sector, and then satisfy the following two
relations among the observables:
\begin{eqnarray}
  |V_{cb}| \,=\, \sqrt{\frac{m_c}{m_t}} \ , \qquad\quad
 | V_{us}V_{cb}| \,=\, |V_{ub}| \ ,
\end{eqnarray}
with which we would have some discrepancy between the prediction
and the experimental data.

\subsubsection{$\bf (Q_2,Q_3)+Q_1$, $\;\;\bf (u_2,u_3)+u_1$, 
$\;\;\bf (d_1,d_2)+d_3$} 
\label{case4}

The last case is the $S_3$ charges; $(Q_2,Q_3)$, $(u_2,u_3)$ 
and $(d_1,d_2)$ are the doublets, which assignment is also consistent
with $SU(5)$ unification. Note that this type of assignment differs from
the third case above, only for the doublet constitution of
right-handed down quarks in the first and second
generations. Therefore the third and this fourth cases are physically
equivalent if all parameter spaces were taken into account. That is,
these two cases correspond to two separate parameter regions in a
single theory, both of which regions are phenomenologically
viable. The mass texture forms now become
\begin{equation}
  M_u \,=\, \left(
  \begin{array}{ccc}
    & b_u & \\ d_u & & f_u \\ & -f_u & i_u
  \end{array}\right)  \ , \qquad\quad
  M_d \,=\, \left(
  \begin{array}{ccc}
    & b_d & \\ d_d & e_d & \\ -e_d & & i_d
  \end{array}\right) \ ,
  \label{matrix4}
\end{equation}
which give the masses and mixing angles in the first order
approximation:
\begin{eqnarray}
  & m_u \,=\, -\frac{b_ud_ui_u}{f_u^2} \ ,\qquad\quad 
  m_c \,=\, \frac{f_u^2}{i_u}\ ,  \qquad\quad 
  m_t \,=\, i_u \ , \\
  & m_d \,=\, -\frac{b_dd_d}{e_d} \ , \qquad\quad 
  m_s \,=\, e_d \ , \qquad\quad 
  m_b \,=\, i_d \ , \\
  & V_{us} \,=\, -\frac{b_ui_u}{f_u^2}+\frac{b_d}{e_d} \ , \qquad\quad 
  V_{cb} \,=\, -\frac{f_u}{i_u} \ , \qquad\quad
  V_{ub} \,=\, \frac{b_u}{f_u} \ .
  \label{vus4}
\end{eqnarray}
There are 8 free parameters for 9 observables, and we have one
predictive relation
\begin{equation}
  |V_{cb}| \,=\, \sqrt{\frac{m_c}{m_t}} \ .
  \label{rel4}
\end{equation}
This relation is known to well fit the experimentally observed values
and has already been discussed in other theoretical
frameworks~\cite{WY,Vcb}. However the precise numerical estimation
indicates that the equation (\ref{rel4}) is not exactly satisfied with
the observational data: $|V_{cb}|=0.039-0.044$ 
and $\sqrt{m_c/m_t}=0.057-0.064$~\cite{PDG}. Some remedies can
easily be found. The discrepancy is removed with radiative
corrections, for instance, the renormalization-group effects on masses
and mixing angles. A naive and probable source of such effects is the
large top-quark Yukawa coupling. 
(The strong gauge coupling does not
affect mass ratios and generation mixing due to its flavor
universality.) \ 
For example, in the MSSM we obtain the
renormalization-group equation
\begin{equation}
  \frac{d\ln\Big(|V_{cb}|\big/\sqrt{m_c/m_t}\Big)}{d\ln\mu} \,=\, 
  \frac{1}{32\pi^2}\left(y_t^2-y_b^2\right).
\end{equation}
where $\mu$ is the renormalization scale and $y_{t},y_{b}$ are the top 
and bottom Yukawa couplings, respectively.
The positive coefficient of the top-Yukawa contribution implies that
the discrepancy is indeed reduced in lower energy regime. It is
however noted that in the present model the flavor symmetry is
supposed to be broken at low energy such as the electroweak scale and
hence the scale dependence might not be enough to make the relation
(\ref{rel4}) fulfilled.

Another possible source of radiative corrections comes from
supersymmetry-breaking parameters. The scalar fermions propagate in
the loops and the chirality is flipped via holomorphic or
non-holomorphic couplings of scalar quarks. In both of these cases,
the corrections do not change zero structure of mass textures but
could modify nonzero matrix elements, depending on
supersymmetry-breaking mass parameters of scalar quarks. For example,
if the $(M_u)_{23,32}$ elements receive such corrections, the 
relation (\ref{rel4}) is modified and the theory would become viable
in light of the current experimental data. We leave a detailed study
of model dependence on supersymmetry-breaking parameters of matter
fields to future investigations.

In the present analysis, we have neglected CP-violating phases.
All the mass matrix elements in this subsection can be taken to 
be real by quark phase redefinitions.
This fact implies that the mass matrices are diagonalized by real
orthogonal matrices up to overall phase rotations. In this case
the observable CP phase is induced by the presence of different
phase rotations between the up and down left-handed quarks.
This is the case for the quark mass matrices in this subsection.
The prediction of quark sector CP violation is consistent with 
the experimental data as long as the mixing angles are properly
reproduced.

\subsection{Lepton mass textures}

Under the standard model gauge symmetry, the flavor charge assignment 
and Yukawa couplings of the lepton sector are free from those of the
quark sector. One could explore the patterns of $S_3$ charges for
lepton fields, just as in the previous analysis for the quark sector,
so that various phenomenological constraints are satisfied. However an
attractive way to determine lepton flavor charges is to promote the
theory to be embedded into grand unification, where lepton and quark
multiplets are unified and have the same flavor charges. Along this
line of thought, the cases 2, 3, and 4 in the previous section are the
candidates to be considered. It is however easily found that the cases
2 and 3 are not reconciled with the experimental data of lepton
sector, even if one includes additional Higgs fields in
higher-dimensional representations of unified gauge symmetry. Thus we
find the unique solution, the case 4 (in Section~\ref{case4}), 
for $S_3$ flavor charges of lepton fields which would be compatible
both with unification hypothesis and the observed data. It may be
interesting to remind that, only from the analysis of quark sector
performed in the previous sections, the case 4 has been
phenomenologically singled out.

Let us first consider the charged lepton sector. We assume 
here $SU(5)$ grand unification because the quark flavor charges in the
case 4 are consistent with it. The left-handed charged 
leptons $L_i$ are combined with the right-handed down quarks into
unified gauge multiplets (anti quintuplets), and 
therefore $(L_1,L_2)$ consists of an $S_3$ doublet and $L_3$ a
singlet. Similarly, $(e_2,e_3)$ transforms as an $S_3$ doublet 
and $e_1$ as a singlet. The charged lepton mass matrix $M_e$ is given
by the transpose of that of down quarks $M_d$ in (\ref{matrix4}) and
becomes at low energy
\begin{eqnarray}
  M_e \,\simeq\, r\left(
  \begin{array}{ccc}
    & d_d & 3 e_d \\[1mm]
    b_d & -3 e_d & \\[1mm]
    & & i_d
  \end{array} \right) \ , 
  \label{Me}
\end{eqnarray}
where we have included a group-theoretical factor `$-3$'~\cite{GJ} in
front of the element $e_d$. Such a factor originates from a Yukawa
coupling to higher-dimensional Higgs field [e.g.\ 45-plet of $SU(5)$] 
\footnote{The representation of Higgs fields under unified gauge
symmetry is independent of the potential analysis at the electroweak
scale given in the previous section.}
and is known to make the mass eigenvalues of charged
leptons well fitted to the observed values if one takes account of
renormalization-group effects on the down quark Yukawa couplings from
the strong $SU(3)$ gauge sector [that has been effectively included as
the factor $r$ in (\ref{Me})]. The mass 
matrices $M_d$ (\ref{matrix4}) and $M_e$ (\ref{Me}) are found to
satisfy the relations
\begin{equation}
  \frac{3m_e}{m_d} \,=\, \frac{m_\mu}{3m_s} \,=\, \frac{m_\tau}{m_b},
\end{equation}
which lead to a better explanation for the mass eigenvalues than that
without the group-theoretical factor. We have found that all the other
patterns to include factors $-3$ are not consistent with the
observation. Therefore the charged lepton mass texture (\ref{Me}) is
the unique simplest solution with our vacuum aligned scheme 
in $SU(5)$ grand unification with $S_3$ flavor symmetry.

The mixing matrix which rotates the left-handed charged leptons to
diagonalize $M_e$ is 
\begin{eqnarray}
  U_e \,=\, \left(
  \begin{array}{ccc}
    1 & \frac{d_d}{3e_d} & \frac{3e_d}{i_d}\\[1mm]
    -\frac{d_d}{3e_d} & 1 & - \frac{d_d}{i_d}\\[1mm]
    -\frac{3e_d}{i_d} & 0 & 1
  \end{array} \right) \ , 
  \label{Ue}
\end{eqnarray}
in the first order approximation. It is found from this expression
that $|(U_e)_{13}|=m_\mu/m_\tau\simeq1/17$ and $|(U_e)_{23}|=
|(U_e)_{12}(U_e)_{13}|\ll m_\mu/m_\tau$. In particular, the latter
means that the observed large lepton mixing between the second and
third generations must come from the neutrino sector. As 
for $(U_e)_{12}$, a naive upper bound is obtained from the
experimental upper bound on the 1-3 lepton 
mixing $V_{e3}$; $\;|(U_e)_{12}|<\sqrt{2}(V_{e3})_{\rm max}$, 
where $V_{\alpha i}$ ($\alpha=e,\mu,\tau$ and $i=1,2,3$) is the
observable lepton mixing matrix~\cite{MNS}. Interestingly, a lower
bound on the charged lepton contribution to the 1-2 lepton 
mixing $|(U_e)_{12}|$ can be expressed in terms of the observables by
imagining that there should not be any fine tuning of parameters in
reproducing $V_{us}$ in (\ref{vus4}). That 
gives $|(U_e)_{12}|>3(m_e/m_\mu)/V_{us}\sim1/16$.

We introduce the three generation right-handed neutrinos and utilize
the seesaw mechanism~\cite{seesaw} to obtain tiny neutrino masses. The
neutrino Dirac mass texture $M_\nu$ in the vacuum of theory and the
right-handed neutrino bare Majorana mass matrix $M_R$ are respectively
read from the flavor symmetry invariance and given by
\begin{eqnarray}
  M_\nu \,=\, \left(
  \begin{matrix}
    & b_\nu & c_\nu \\
    -b_\nu & e_\nu & \\
    g_\nu & &
  \end{matrix} \right ), \qquad
  M_R \,=\, \left(
  \begin{matrix}
    & M_1 & \\
    M_1 & & \\
    & & M_2
  \end{matrix} \right).
\end{eqnarray}
Here we have taken the $S_3$ charge of right-handed neutrinos 
as $(\nu_1,\nu_2)+\nu_3$, that is, the first two generations make a
doublet. It is however noticed that the charge assignment 
of $\nu_i$ (equivalently the label changing effect of $\nu_i$) is
completely irrelevant to low-energy physics and the generation
structure of light neutrino mass matrix is determined only by the
flavor charge of left-handed leptons $L_i$. After integrating out
heavy modes, the light neutrino mass 
matrix $M_L=-M_\nu M_R^{-1}M_\nu^T$ becomes
\begin{equation}
  M_L \,=\, \frac{b_\nu^2}{M_1}\left(
  \begin{array}{ccc}
    -z & 1 & -x \\[1mm]
    1 & 2y & -xy \\[1mm]
    -x & -xy & 0 
  \end{array} \right) \ ,
\end{equation}
\begin{equation}
  x \,=\, \frac{g_\nu}{b_\nu} \ , \qquad
  y \,=\, \frac{e_\nu}{b_\nu} \ , \qquad
  z \,=\, \frac{c_\nu^2}{b_\nu^2}\frac{M_1}{M_2} \ .
\end{equation}
Taking into account that the charged lepton sector has small
generation mixing, it is found that this form of neutrino mass matrix
is suitable for large generation mixing with the inverted mass
hierarchy of light neutrinos: $m_2\simeq m_1\gg m_3$. The 2-3 large
lepton mixing is controlled by $x\sim{\cal O}(1)$. The observed mass
squared differences imply the traceless 
condition $2y\simeq z$ ($|y|,|z|\ll1$).
 In the limit $y,z\to0$, the
1-2 and 2-3 neutrino generation mixings are 
maximal ($\theta_{12}^\nu=\theta_{23}^\nu=\pi/4$) and the 1-3 angle
becomes zero ($\theta_{13}^\nu=0$). Including finite effects 
of $y$ and $z$, we obtain the first order expressions:
\begin{equation}
  \tan^2\theta_{12}^\nu \,=\, 1-\frac{2y+z}{\sqrt{1+x^2}} \ , \qquad
  \tan\theta_{23}^\nu \,=\, x \ , \qquad
  \tan\theta_{13}^\nu \,=\, \frac{-xy}{\sqrt{1+x^2}} \ ,
\end{equation}
\begin{equation}
  m_1 = \frac{b^2_\nu}{M_2}\left(\sqrt{1+x^2}-y+\frac{z}{2}\right) \ , 
  \qquad
  m_2 = \frac{b^2_\nu}{M_2}\left(\sqrt{1+x^2}+y-\frac{z}{2}\right) \ ,
  \qquad
  m_3 = 0 \ .
\end{equation}
If the small mixing contributions from $M_e$ are neglected, there is a
prediction which relates the observables in neutrino oscillation
experiments:
\begin{equation}
  \frac{2\tan\theta_{13}^\nu}{\tan\theta_{23}^\nu}
  +\tan\theta_{12}^\nu-1 \,=\, \frac{m_2^2-m_1^2}{4(m_2^2-m_3^2)} \ .
  \label{nurel}
\end{equation}
This is certainly consistent with the current experimental
data. Finally, including the charged lepton contribution (\ref{Ue}),
we obtain
\begin{eqnarray}
  V_{e2} &=& \sin\theta_{12}^\nu
  -(U_e)_{12}\cos\theta_{12}^\nu\cos\theta_{23}^\nu
  +(U_e)_{13}\cos\theta_{12}^\nu \sin\theta_{23}^\nu  \ , \\
  V_{\mu 3} &=& \sin\theta_{23}^\nu \ , \\
  V_{e3} &=& \sin\theta_{13}^\nu -(U_e)_{12}\sin\theta_{23}^\nu 
  -(U_e)_{13}\cos\theta_{23}^\nu \ ,
\end{eqnarray}
where $\theta_{13}^\nu\ll 1$ has been used. 
It is found 
that $V_{e3}$ has a lower bound; $|V_{e3}|\geq 0.04$, which is derived
 by taking account of the experimental data, $V_{e2}$, $V_{\mu 3}$,
$\Delta m^2_{12}$ and $\Delta m^2_{23}$ \cite{PDG} with 
the relation (\ref{nurel}) and the constraints 
on $(U_e)_{12,13}$ discussed above. 
That will be tested in future
experiments such as the double Chooz~\cite{doublechooz}.

\medskip

In Section 5, we have shown that our scheme for generating zero
texture forms with vacuum alignment is applied to mass matrices of
quarks and leptons. Exhausting the patterns of matter flavor charges,
we have found the highly limited numbers of charge assignments are
phenomenologically viable. In particular, our scheme is also
consistent with $SU(5)$ grand unification, and the uniquely determined 
matter flavor charges predict typical low-energy phenomenology such as
the relations among the observables, independently of model parameters.

\bigskip
\section{Higgs mass spectrum and $S_3$ breaking terms}

The analysis in the previous section has shown that the 
vacuum (\ref{vev0}) is phenomenologically interesting if applied to
the quark and lepton mass textures. However the model has a problem
that some massless bosons generally appear at any vacuum where more
than one sets of Higgs bosons have non-vanishing VEVs. This general
feature comes from the fact that the Higgs scalar potential given in
Section 3 has an enhanced global symmetry $SU(2)\times U(1)^2$ and
leads to massless Nambu-Goldstone bosons in the electroweak broken
phases. It is therefore reasonable to (softly) break the flavor
symmetry within the scalar potential so as not to enhance the global
symmetry. In this paper we introduce supersymmetry-breaking soft terms
which violate $S_3$ flavor symmetry. We find that the following simple
set of breaking terms is appropriate for the present purpose:
\begin{equation}
  V_{\not S_3} \,=\, b_{SD}H_{uS}H_{d2} +b'_{SD}H_{u1}H_{dS} 
  +b_{AD}H_{uA}H_{d1} +b'_{AD}H_{u2}H_{dA} +{\rm h.c.}\ ,
\label{S3break}
\end{equation}
There are several reasons why these terms are chosen: (i) The global
symmetry of total Higgs potential is broken down to $U(1)_X$ under
which $H_{uS}$ and $H_{u1}$ ($H_{dS}$ and $H_{d2}$) are positively
(negatively) charged. Note that, combined with 
the $U(1)_Y$ hypercharge symmetry, $U(1)_X$ is converted to another
global $U(1)$ where $H_{uA}$ and $H_{d1}$ ($H_{dA}$ and $H_{u2}$) are
positively (negatively) charged (see Table~\ref{U(1)}).
\begin{table}[t]
\begin{center}
\begin{tabular}{|c|cccccccc|} \hline
& $H_{uS}$ & $H_{dS}$ & $H_{uA}$ & $H_{dA}$ & $H_{u1}$ & $H_{u2}$ &
$H_{d1}$ & $H_{d2}$ \\ \hline
$U(1)_X$ & +1 & $-1$ & 0 & 0 & +1 & 0 & 0 & $-1$ \\ \hline
$U(1)_Y$ & +1 & $-1$ & +1 & $-1$ & +1 & $+1$ & $-1$ & $-1$ \\ \hline
\end{tabular}
\end{center}
\caption{The $U(1)$ symmetries of the Higgs scalar potential with 
the $S_3$ breaking terms (\ref{S3break}).\bigskip}
\label{U(1)}
\end{table}
As discussed in Section~\ref{DHiggs}, there are four sets of up and
down type Higgs fields in the original potential without
flavor-breaking terms, which potential therefore has 
four $U(1)$ invariances. Each set of ($H_u,H_d$) are vector-like
fields under these $U(1)$s. Including the soft terms (\ref{S3break})
economically breaks two of them, and hence $U(1)_{X,Y}$ in
Table~\ref{U(1)} remain intact. It is important to notice that, in the
electroweak broken vacuum (\ref{vev0}), $U(1)_X$ is unbroken and the
Nambu-Goldstone boson of $U(1)_Y$ is partly absorbed by the massive
gauge boson $Z$ and becomes unphysical. Therefore no physical
Nambu-Goldstone fields are associated with symmetry breaking. We
numerically checked by examining the neutral Higgs mass matrix that 
there appear no massless scalar bosons. (ii) The second reason is
that, when including the breaking terms (\ref{S3break}), the
stationary conditions can still be solved without imposing any tuning
of model parameters. (iii) Flavor breaking effects in
supersymmetry-breaking holomorphic mass terms do not propagate to
other sectors. Such a favorable property for the model simplicity does
not hold in cases that flavor symmetry violation resides in other
supersymmetry-breaking sectors.

The model also predicts a light Higgs boson exactly parallel to the
MSSM case. In order to find an approximate form of the Higgs mass
spectrum, we take a simplifying assumption:
\begin{eqnarray}
  & \left|\mu_x\right|^2 +m^2_{ux} \,=\, 
  \left|\mu_x\right|^2 +m^2_{dx} \,\equiv\, \bar m^2, \qquad
  b_x \,\equiv\, \bar b, \qquad  (x=S,A,D) \\
  & b_{SD}=b'_{SD}=b_{AD}=b'_{AD} \,\equiv\, \bar b'\ ,
\end{eqnarray}
with a hierarchy ${\bar m}^2,\,|\bar b|,\,|\bar b'|\,\gg\,v^2
\equiv v^2_{u2}+v^2_{d1}+v^2_{uA}+v^2_{dA}$. In this case, we can
analytically write down the mass matrices of Higgs fields. In
particular, for the real parts of neutral Higgs bosons, the mass
matrix is given by
\begin{equation}
M_H^2 =
\begin{array}{c}
  h^0_{u1} \\ h^0_{u2} \\ h^0_{uA} \\ h^0_{uS} \\ h^0_{d2} \\
  h^0_{d1} \\ h^0_{dA} \\ h^0_{dS}
\end{array}
\left(\begin{array}{cccccccc}
 \bar m^2 & & & & -\bar b & & & -\bar b' \\
 & \bar b\frac{v_{d1}}{v_{u2}} +\bar b'\frac{v_{dA}}{v_{u2}} 
 & & & & -\bar b & -\bar b' & \\
 & & \bar b\frac{v_{dA}}{v_{uA}} +\bar b'\frac{v_{d1}}{v_{uA}} 
 & & & -\bar b' & -\bar b & \\
 & & & \bar m^2 & -\bar b' & & & -\bar b \\
 -\bar b & & & -\bar b' & \bar m^2 & & & \\
 & -\bar b & -\bar b' & & & \bar b\frac{v_{u2}}{v_{d1}} 
 +\bar b'\frac{v_{uA}}{v_{d1}} & & \\ 
 & -\bar b' & -\bar b & & & & \bar b\frac{v_{uA}}{v_{dA}} 
 +\bar b'\frac{v_{u2}}{v_{dA}} & \\
 -\bar b' & & & -\bar b & & & & \bar m^2 
\end{array}\right) ,
\end{equation}
where $h^0_x$ ($x=u1,u2,\cdots$) are the neutral components of Higgs
bosons. Diagonalizing this matrix, we obtain the neutral Higgs masses
squared:
\begin{equation}
  \{M^2_{h^0},\ M^2_{H^0_1},\ M^2_{H^0_2},\ M^2_{H^0_3},\ M^2_{H^0_4}\}
  \,=\, \{ {\cal O}(v^2),\ \bar m^2-\bar b -\bar b',\ 
  \bar m^2 +\bar b -\bar b',\ \bar m^2 -\bar b +\bar b',\ 
  \bar m^2 +\bar b +\bar b'\},
\end{equation}
and the other three mass eigenvalues squared are written in rather 
complicated expressions, but of the order 
of ${\cal O}(\bar m^2,\bar b,\bar b')$. We therefore have a light mode
with a weak scale mass ${\cal O}(v)$ and its eigenfunction is
explicitly given by
\begin{equation}
  h^0 \,=\, \frac{1}{v}\left( v_{uA} h^0_{uA} +v_{dA} h^0_{dA} 
    +v_{u2}h^0_{u2} +v_{d1}h^0_{d1} \right).
\end{equation}
This light Higgs mode receives a sizable radiative correction from
large top Yukawa coupling similar to the MSSM, and would be made heavy
enough to satisfy the lower bound from the LEP experiment~\cite{PDG}.

\bigskip
\section{Tree-level FCNC}

Since there are multiple electroweak doublet Higgses which couple to
matter fields, flavor-changing processes are mediated at classical
level by these Higgs fields. In the previous section, we show that all
but one Higgs bosons have masses of the order of
supersymmetry-breaking parameters. Therefore the experimental
(un)observations of FCNC rare events would lead to a bound on the
supersymmetry breaking scale. Among various experimental constraints,
we find the most important constraint comes from the neutral K meson
mixing, which gives a lower bound on heavy Higgs masses being larger
than a few TeV\@.

\begin{figure}[t]
\centerline{\includegraphics[scale=0.8]{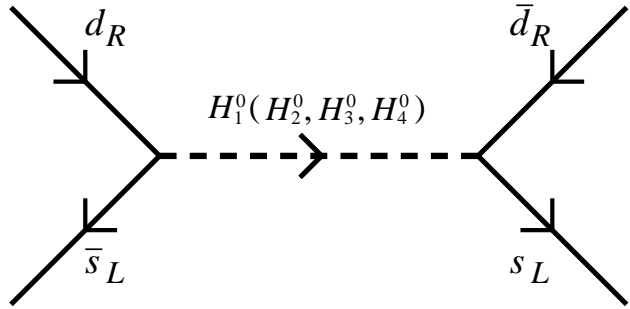}}
\caption{A Higgs-mediated tree-level FCNC process for the K meson
system.\bigskip}
\label{FCNC-K}
\end{figure}

The down quark mass matrix in Eq.~(\ref{matrix4}) is diagonalized as
\begin{equation}
  U_L^\dagger M_dU_R^{} \,=\, 
  \left(\begin{array}{ccc}
    m_d  & & \\
    & m_s & \\
    & & m_b
  \end{array}\right)\ ,
\end{equation}
\begin{equation}
  U_L \,\simeq\, \left(
  \begin{array}{ccc}
    1 & \frac{b_d}{e_d} & 0 \\
    -\frac{b_d}{e_d} & 1 & -\frac{e_dd_d}{i^2_d} \\
    -\frac{b_dd_d}{i^2_d} & \frac{e_dd_d}{i^2_d} & 1 
  \end{array} \right), \qquad\quad
  U_R \simeq \left(
  \begin{array}{ccc}
    1 & \frac{d_d}{e_d} & -\frac{e_d}{i_d} \\
    -\frac{d_d}{e_d} & 1 & 0 \\
    \frac{e_d}{i_d} & \frac{d_d}{i_d} & 1 
  \end{array} \right)\ .
\end{equation}
Here and hereafter in this section, we simply 
take $b_d\simeq d_d$. The other parameters are fixed by the mass
eigenvalues. Tree-level FCNC is mediated by Higgs fields like the
diagram shown in Fig.~\ref{FCNC-K}.
For the heavy mass eigenstates $H^0_i$ ($i=1,2,3,4$), the 
tree-level $K_L$-$K_S$ mass difference $\Delta m_K^{\rm tree}$ is
given by the matrix element of the effective Hamiltonian between K
mesons~\cite{FCNC}, which is analytically evaluated from the above
mixing matrices in the present case as
\begin{eqnarray}
  \Delta m_K^{\rm tree} &=& 2\,{\rm Re}\,
  \big\langle K^0\big|H_{\rm eff}\big|\bar K^0\big\rangle \nonumber \\
  &\sim& \frac{m_b^2m_Kf_K^2}{6v_{d1}^2 M_H^2}\left[\,
    \left(\frac{m_K}{m_s+m_d}\right)^2 \left(\frac{m_s}{m_b}\right)^2
    -\eta^2\left\{\left(\frac{m_K}{m_s+m_d}\right)^2
      +\frac{1}{2}\right\} \,\right]\ ,
\end{eqnarray}
where $m_K$ and $f_K$ are the mass and decay constant of K meson 
and $M_H$ is 
an average of the Higgs 
masses, 
 $ 1 / M_H^{2} = \frac{1}{4}\big( 1 / M_{H^0_1}^{2} + 1 / M_{H^0_2}^{2} + 
1 / M_{H^0_3}^{2} + 1 / M_{H^0_4}^{2} \big)$. 
The parameter $\eta$ contains the
down-type quark Yukawa couplings which do not contribute to mass terms
and explicitly given by
\begin{equation}
  \eta \,=\, \frac{(y_d^S)_{22}b_dv_{d1}}{m_b^2}
  -\frac{(y_d^S)_{13}d_dv_{d1}}{m_sm_b}\ ,
\end{equation}
where $y_d^S$ denotes the matrix of Yukawa coupling of down-type
quarks to the $S_3$ singlet Higgs boson. The $S_3$ singlet 
Higgs $H_S$ has a vanishing VEV in the present vacuum, and 
therefore $\eta$ is regarded as a free parameter. For 
example, $\eta=0$ if one does not include $H_S$ in the theory. The
other Higgs fields also contribute to the K meson mixing in a similar
order.

\begin{figure}[t]
\centerline{\includegraphics[scale=0.8]{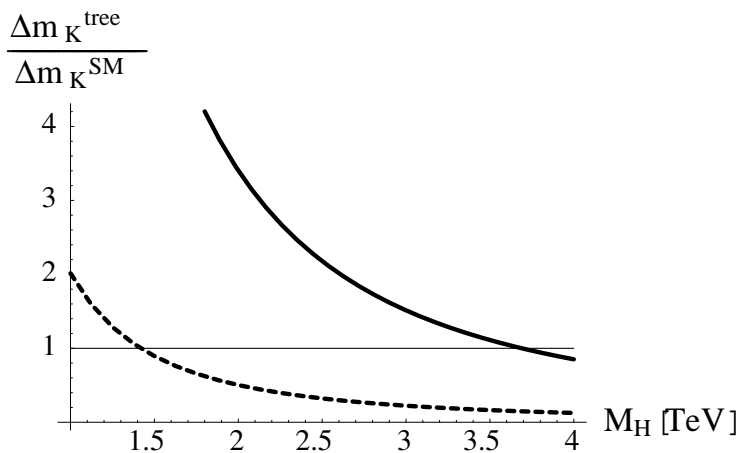}}
\caption{$\Delta m_K^{\rm tree}/\Delta m_K^{\rm SM}$ as the function
of Higgs mass parameter $M_H$. The solid and dashed lines correspond
to $\eta=0$ and $0.03$, respectively.\bigskip}
\label{ratio}
\end{figure}

In order to estimate the bound on supersymmetry breaking scale, we
calculate the ratio of the exact numerical 
value $\Delta m_K^{\rm tree}$ to the standard 
contribution $\Delta m_K^{\rm SM}$. In Fig.~\ref{ratio}, we show the
ratio versus the averaged Higgs mass $M_H$ for the cases 
of $\eta=0$ and $\eta=0.03$ as typical values.
It is found from the figure that heavy Higgs masses, in turn the
supersymmetry breaking masses, are bounded from below so as to
suppress the extra Higgs contribution compared with the standard model
one, which bound is roughly given by
\begin{equation}
  M_H \,\geq\, \left\{
  \begin{array}{rcl} 
    3.8 {\rm ~TeV} && (\eta=0) \\[2mm]
    1.4 {\rm ~TeV} && (\eta=0.03) \\
  \end{array} \right.,
\label{bound}
\end{equation}
where we have used the experimental data $m_K=490$~MeV, $f_K=160$~MeV
and taken $v_{d1}=100$~GeV as a typical electroweak scale. Other
flavor-changing rare processes are also suppressed for such heavy
Higgs fields of few TeVs.
For example, the $\mu\rightarrow e+\gamma$ process is
given in terms  of extra Higgs masses and lepton Yukawa couplings
 \cite{mue}.  The branching ratio is found to be suppressed  enough 
under the constraint of eq.({\ref{bound}).

\bigskip
\section{Summary}

In this paper we have discussed the structure of Higgs potential and
fermion mass matrices in supersymmetric models with $S_3$ flavor
symmetry. The electroweak doublet Higgs fields belong to non-trivial
representations of the flavor symmetry and lead to restricted forms of
vacuum structure. We have examined possible zero elements (textures)
of quark and lepton mass matrices. Our approach to have mass textures
is rather different from previous ones in a sense that flavor symmetry
does not forbid coupling constants of (effective) Yukawa operators,
but leads to mass texture zeros by ensuring some of Higgs fields have
vanishing expectation values in the vacuum of the theory. The
vanishing mass matrix elements are obtained dynamically in the vacuum
without any exact tuning of model parameters, and their positions are
controlled by flavor symmetry. An interesting point is that, due to
the flavor group structure, the up and down quark mass matrices are
automatically made different, which lead to non-vanishing generation
mixing. We have exhausted the patterns of flavor symmetry charges of
matter fields and found that the simplest, viable forms of mass
matrices including the neutrino sector are uniquely determined. In
particular the lepton mixing $V_{e3}$ is predicted within the range
that will be tested in near future experiments.

We have also discussed the physical mass spectrum of Higgs bosons and
its phenomenological implication. The flavor symmetry is softly broken
by a certain class of holomorphic supersymmetry-breaking mass terms of
Higgs bosons. That neither destabilizes the desired vacuum nor
introduces any fine tuning of coupling constants. Given these flavor
breaking terms, it is found that all Higgs bosons, except for the
lightest one, can be made heavier than a few TeV enough to satisfy the
experimental bounds such as that from the neutral K meson mixing. 
Detail phenomenological  analyses including CP violation and FCNC
will be presented elsewhere.

\vspace{1cm}
We would like to thank J. Kubo, T. Kobayashi and H. Nakano for helpful 
discussions. This work was supported in part by scientific grants from the 
Ministry 
of Education, Science, Sports, and Culture of Japan (No.~17540243 and
17740150), and grant-in-aid for the scientific research on priority
area (\#441) "Progress in elementary particle physics of the 21st
century through discoveries of Higgs boson and 
supersymmetry" (No.~16081209).


\end{document}